\documentclass[10pt]{IEEEtran}
\usepackage{amssymb, algorithm, algorithmic, amsmath, graphicx, psfig}

\newcommand{\bl}{\boldsymbol}
\newcommand{\ml}{\mathcal}
\newcommand{\diff}{{d\over dt}}
\newcommand{\tsf}{\textsf}
\newtheorem{lemma}{Lemma}
\newtheorem{theorem}{Theorem}
\newtheorem{corollary}{Corollary}

\title{Maximal Scheduling in Wireless Networks with Priorities}
\author{Qiao~Li,~\IEEEmembership{Student~Member,~IEEE,} and Rohit~Negi,~\IEEEmembership{Member,~IEEE} \thanks{Parts of this work were presented at IEEE Globecom 2008 \cite{li08} and IEEE Infocom 2009 \cite{li09}. Qiao Li and Rohit Negi are with the Department of Electrical and Computer Engineering, Carnegie Mellon University, Pittsburgh, PA 15213 USA (email: qiaoli@cmu.edu, negi@ece.cmu.edu).}}

\begin{document}

\maketitle

\begin{abstract}
  We consider a general class of low complexity distributed scheduling algorithms in wireless networks, \emph{maximal scheduling with priorities}, where a maximal set of transmitting links in each time slot are selected according to certain pre-specified static priorities. The proposed scheduling scheme is simple, which is easily amendable for distributed implementation in practice, such as using inter-frame space (IFS) parameters under the ubiquitous 802.11 protocols. To obtain throughput guarantees, we first analyze the case of maximal scheduling with a fixed priority vector, and formulate a lower bound on its stability region and scheduling efficiency. We further propose a low complexity priority assignment algorithm, which can stabilize \emph{any arrival rate} that is in the union of the lower bound regions of all priorities. The stability result is proved using fluid limits, and can be applied to very general stochastic arrival processes. Finally, the performance of the proposed prioritized maximal scheduling scheme is verified by simulation results.
\end{abstract}


\maketitle

\section{Introduction}
\label{sec_intro}

\IEEEPARstart{T}{he} design of efficient scheduling in wireless networks has attracted much attention over the past few years (e.g. \cite{hajek88, tassiulas92, chaporkar08, jiang10, li11}). As the \emph{core subproblem} in the cross-layer optimization for wireless networks \cite{lin06}, the MAC-layer scheduling plays a key role in achieving efficient and fair utilization of the wireless network resources. On the other hand, the MAC-layer scheduling is very challenging, due to the complicated conflicting relationships between transmitting links, due to the fundamental \emph{broadcast nature} of wireless communications. That is, the transmission of any link will be received by any unintended receiver, which can be arbitrarily located in the network, as \emph{interference}, and thereby impairing its own communication quality.

Despite the numerous efforts made in the past, optimal MAC-layer scheduling in wireless networks is still hard to achieve. For example, the popular optimal max-weight scheduling by Tassiulas and Ephremides \cite{tassiulas92} is hard to implement even in a centralized manner. This is because the algorithm requires solving a max-weight independent set (MWIS) problem \emph{in each time slot}, which is well-known to be NP-hard for wireless networks under general interference constraints. To resolve the complexity issue, several attempts were made to achieve optimal distributed scheduling using constant computation per time slot. For example, Tassiulas \cite{tassiulas98} proposed an optimal random `pick-and-compare' scheduling with linear complexity per time slot. Recently, Jiang \emph{et al.} \cite{jiang10} and Ni \emph{et al.} \cite{ni10} proposed CSMA based optimal scheduling schemes, which only requires constant computation per time slot. However, all such algorithms suffer from the large delay (exponential in the size of the network) in the worst case, which is inevitable \cite{shah10}, since, intuitively, it takes an exponential number of time slots for such \emph{amortized} `constant computation' based schedules to converge to an optimal schedule, due to the NP-hardness of the scheduling problem \cite{arikan84}.

\begin{figure*}[!t]
  \begin{center}
    \begin{tabular}{cc}
    \psfig{figure=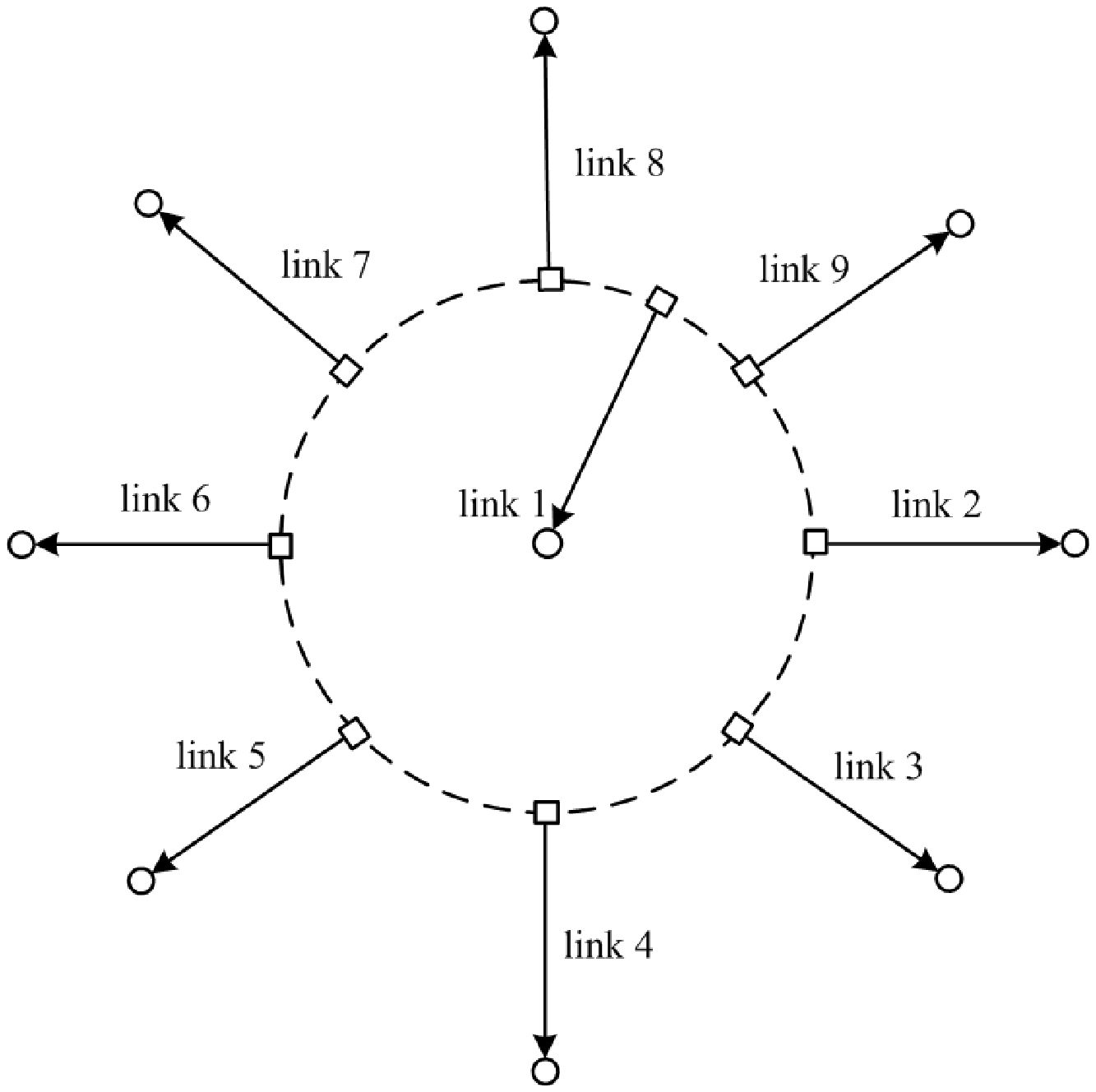,width=3.2in} & \psfig{figure=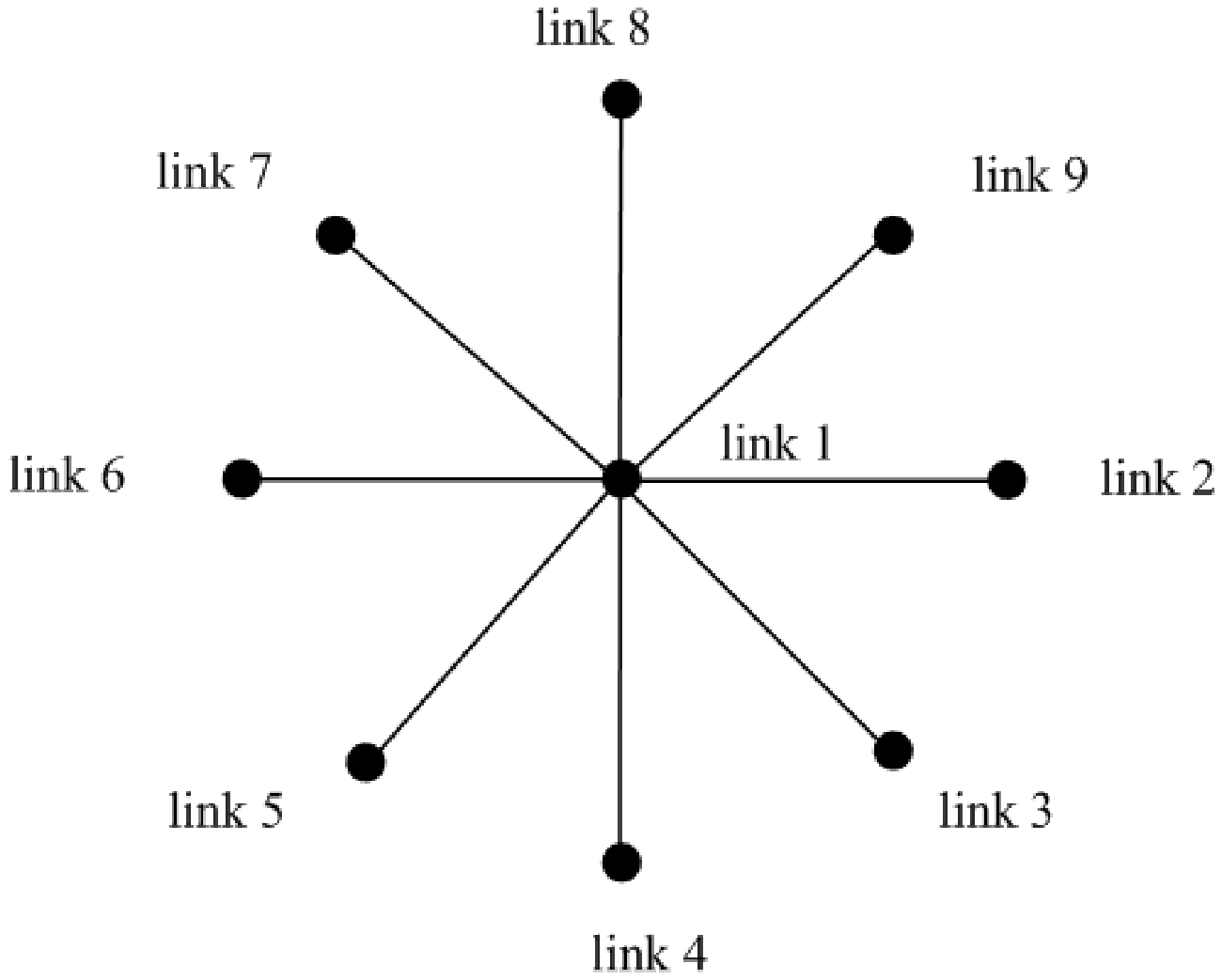,width=3.2in}\\
    (a) & (b)
    \end{tabular}
  \end{center}
  \caption{(a) is a wireless network of $9$ links, where square nodes denote transmitters and round nodes denote receivers. (b) is its interference graph.}
  \label{fig_star}
\end{figure*}

As such, \emph{distributed suboptimal scheduling}, even if it achieves only a fraction of the maximum throughput region, is still very attractive, due to the low complexity and ease of implementation. Recently, Chaporkar \emph{et al.} \cite{chaporkar08} have shown that a class of simple scheduling policies, \emph{maximal scheduling}, can achieve a guaranteed fraction of the optimal stability region for general wireless networks, which can even be constant for large-scale wireless networks under the ubiquitous 802.11 protocols \cite{chaporkar08}. The scheduling scheme is very simple. Under the popular interference graph model, a maximal scheduler simply chooses a maximal set of backlogged links that form an independent set of the interference graph. A set of links is maximal if it can not be further augmented. The scheduling is otherwise arbitrary. For example, consider the wireless network in Fig. \ref{fig_star} (a) and its interference graph in Fig. \ref{fig_star} (b). The set $\{1\}$ is a \emph{maximal} independent set. The set $\{2, 3, 4, \ldots, 8\}$ is a also maximal, but in addtion, is a \emph{maximum} independent set, since it is the independent set with the largest cardinality. Compared with the optimal scheduling schemes \cite{hajek88, tassiulas92, jiang10, tassiulas98, ni10}, maximal scheduling is very attractive, as it can achieve quite good throughput performance with distributed implementation, with low complexity \cite{gupta09}, or even constant overhead \cite{lin09}.

On the other hand, maximal scheduling suffer from small throughput guarantees under certain topologies, due to the \emph{ad hoc} choice of maximal schedules. One example is the general star-shaped interference graph such as in Fig. \ref{fig_star} (b) with $N$ peripheral links, where it has been shown \cite{chaporkar08} that the worst case maximal scheduler can only achieve $1/N$ of the optimal stability region. In the extreme case, as $N\rightarrow\infty$, the worst maximal scheduling can not achieve \emph{any positive fraction} of the optimal stability region for certain packet arrival processes \cite{chaporkar08}. Thus, in order to improve the throughput performance of maximal scheduling, it is essential to carefully design the scheduling scheme by choosing the schedules carefully according to the network parameters, such as topology and packet arrivals.

In this paper, we improve the performance of maximal scheduling by using \emph{static priorities}, which are chosen based on the network topology and packet arrival rates. During scheduling, the scheduler simply considers the links in a sequence specified by a priority vector $\bl p$, and adds back-logged links to the schedule whenever there is no conflict. It can be easily shown that the resulting independent set is maximal. The scheduling is simple, and is easily amendable for distributed implementation. For example, the ubiquitous 802.11 protocols has already defined a set of Inter Frame Space (IFS) parameters to provide prioritized wireless channel access. A related algorithm is the Longest Queue First (LQF) scheduling (also referred to as the greedy maximal scheduling) \cite{dimakis06}, where the priorities are chosen according to queue lengths in each time slot, so that a link with higher queue length will receive higher priority. Compared to the LQF scheduling, where the priorities can change globally every time slot, the static priority approach requires much smaller overhead, and is therefore easier for distributed implementation.

Simple as the prioritized maximal scheduling scheme is, the assignment of the static priorities is highly nontrivial, one has to search over a set of $N!$ possible priorities for a network with $N$ links. For networks with moderate size, a naive search over all possible priorities simply becomes impossible. As a main contribution of this paper, we show that, somewhat surprisingly, the optimal priority assignment can be achieved \emph{online}, and with only \emph{linear complexity}. The optimality is in the following sense. For maximal scheduling with a priority vector $\bl p$, we associate a tight lower bound stability region $\Lambda_{\bl p}$, and prove that, using fluid limits, the network is stable under $\bl p$ whenever the general stochastic packet arrivals have average rate $\bl \lambda\in \Lambda_{\bl p}$. Now, suppose a packet arrival rate vector $\bl \lambda$ is given. We claim that, as long as $\bl \lambda\in \Lambda_{\tsf{sp}}\triangleq\cup_{\bl p\in\ml P} \Lambda_{\bl p}$, where $\ml P$ is the set of $N!$ priority vectors, our proposed priority assignment algorithm can produce a stabilizing priority vector $\bl p'$, such that $\bl \lambda$ is stable under $\bl p'$. In other words, by obtaining the maximal scheduler based on a carefully chosen priority vector, we can improve the guaranteed throughput region of maximal scheduling from $\Lambda_{\textsf{wc}}\subseteq \cap_{\bl p\in\ml P}\Lambda_{\bl p}$ to $\Lambda_{\textsf{sp}}=\cup_{\bl p\in\ml P} \Lambda_{\bl p}$. Such throughput increase can significantly improve the cross-layer optimization results \cite{lin06} by providing upper layers with larger guaranteed throughput regions.

Based on the obtained stability region $\Lambda_{\tsf{sp}}$, we next analyze its \emph{scheduling efficiency}, which is defined as the largest fraction of the optimal throughput region. We show that the scheduling efficiency can be bounded by $1/\Delta_{\tsf{sp}}$. Here $\Delta_{\tsf{sp}}$ is the \emph{prioritized interference degree} of the network, which is similar to the `interference degree' metric defined in \cite{chaporkar08}. However, the lower bound $1/\Delta_{\tsf{sp}}$ can be much larger than that in \cite{chaporkar08}. For example, it can be shown that the prioritized interference degree of any \emph{acyclic interference graph} is 1, in which case the prioritized maximal scheduling scheme is \emph{globally optimal}. On the other hand, the interference degree of the same network can be arbitrarily large (e.g., $N$ for the star shaped network in Fig. \ref{fig_star}), in which case the scheduling efficiency of the maximal scheduler is close to zero. Finally, the proposed $1/\Delta_{\tsf{sp}}$ bound is the same as the one used for LQF scheduling \cite{joo09}, so that the static priority based scheduling can achieve similar throughput guarantee as the LQF scheduling, but with much simpler design and lower scheduling overhead.

The organization of the rest of this paper is as follows: In Section \ref{sec_system_model} we formulate the system model for prioritized maximal scheduling, and in Section \ref{sec_fixed} we analyze the performance of maximal scheduling assuming a fixed priority vector. Section \ref{sec_assignment} proposes an online priority assignment algorithm and proves its throughput guarantees, Section \ref{sec_simulation} demonstrates simulation results, and finally Section \ref{sec_conclusion} concludes this paper.

\section{System Model}
\label{sec_system_model}

In this section, we describe the system model and formulate the prioritized maximal scheduling problem. We begin with the network model.

\subsection{Network Model}

We consider a single-hop wireless network, whose topology can be modeled as a directed graph $\mathcal{G}=\ml{(V, E)}$, where $\ml{V}$ is the set of user nodes, and $\ml{E}$ is the set of communication links. Fig. \ref{fig_star}(a) shows a wireless network consisting of 9 links. Note that it is not hard to generalize the analysis in this paper to the multi-hop scenarios using standard techniques \cite{chaporkar08, wu07}. In this paper, the focus on single-hop network is mainly for the simplicity of exposition. The interference constraint is modeled by an undirected interference graph $\ml{G}_I=(\ml{V}_I, \ml{E}_I)$, where $\ml{V}_I$ represents the transmitting links (edges) in $\ml{G}$, and $\ml{E}_I$ is the set of pairwise conflicts. Thus, two links $(i, j)\in \ml{E}_I$ if and only if they are not allowed to transmit together, due to the strong interference one may cause upon the other. In any time slot, the set of scheduled links must form an independent set of $\ml G_I$. Fig. \ref{fig_star}(b) illustrated the interference graph corresponding to the wireless network in Fig. \ref{fig_star}(a). The interference graph model is widely adopted for the scheduling problem in many types of wireless networks, such as the Blue-tooth or FH-CDMA networks, where the interference graph is built based on the \emph{node exclusive interference model} \cite{gupta09}, which specifies that a node can either transmit or receive in each time slot, but not both. It can also be applied to the ubiquitous 802.11 networks, where a \emph{two-hop interference model} \cite{lin09} is used, i.e., links within two hops are not allowed to transmit together, due to the exchange of RTS/CTS control messages.

We assume that time is slotted. Each link $i$ is associated with an exogenous stochastic packet source, which is specified by upper layer protocols. The packet arrivals happen only at the end of each time slot. We have the following weak assumptions on the packet arrival processes. First, the number of arrived packets in a time slot is uniformly bounded with probability 1 (w.p.1):
    \begin{equation}
      A_i(t)-A_i(t-1)\leq A_{\max}, \forall i\in\ml V_I, t\geq 0,
           \label{eqn_arr_bdd}
    \end{equation}
    where $A_i(t)$ is the \emph{total} number of packets that have arrived at link $i$ during the first $t$ time slots, and $A_{\max}$ is a large positive constant. Further, we assume that Strong Law of Large Numbers (SLLN) can be applied to the arrival processes:
    \begin{equation}
      \lim_{t\rightarrow\infty}A_i(t)/t=\lambda_i \qquad \text{ with probability 1,} \ \forall i\in \ml{V}_I
      \label{eqn_slln}
    \end{equation}
Note that these assumptions on the arrival processes are quite mild, since the packet arrivals are allowed to be \emph{arbitrarily correlated} across links as well as over time slots. In each time slot, a scheduler chooses an independent set of back-logged links $\bl \sigma(t)$ for transmission. With an abuse of notation, we also denote $\bl \sigma(t)$ as an $N\times 1$ vector of indicator functions for the independent set. That is, $\sigma_i(t)=1$ if link $i$ is in the independent set, otherwise $\sigma_i(t)=0$. Thus, we can write the total departures until the end of slot $t$ in a vector form as $\bl D(t)=\sum_{\tau=1}^t \bl \sigma(\tau)$. Finally, the queueing dynamics of the network is as follows:
\begin{equation}
  \bl Q(t)=\bl Q(0)+\bl A(t)-\bl D(t)
\end{equation}
where $\bl Q(t)$ is the queue length vector at time slot $t$.

\subsection{Maximal Scheduling with Static Priorities}

As described in Section \ref{sec_intro}, we propose to use a priority vector $\bl p$ to assist maximal scheduling. Here, $\bl p$ is a permutation of $(1, 2, \ldots, N)^T$, where $p_i$ is the priority of link $i$. We say that link $i$ has higher priority than link $j$ if $p_i<p_j$. Thus, the link $i$ with $p_i=1$ has the highest priority, while the link $j$ with $p_j=N$ has the lowest priority. Given $\bl p$, the prioritized maximal scheduler computes the schedule by considering the links sequentially, from the highest priority `$1$' to the lowest priority `$N$', and adds each back-logged link to the schedule if none of its higher priority neighbors have already been scheduled when it is considered. Denote $\ml S_i^{\bl p}$ as the set of higher-priority neighbors of link $i$ according to $\bl p$. The following is a key property for proving the throughput guarantee of the scheduling scheme:

\begin{lemma}\label{lem_pms}
In any time slot, for any back-logged link $i$, a maximal scheduler with priority $\bl p$ will schedule at least one departure among the links $\{i\}\cup \ml{S}_i^{\bl p}$.
\end{lemma}

\begin{IEEEproof}
  See Appendix \ref{apdx_pms}.
\end{IEEEproof}

\subsection{Performance Metrics}

\subsubsection{Stability Region}

The throughput performance of a scheduler can be represented by its \emph{stability region}, which is the set of stable arrival rates under the scheduler. In this paper, we are interested in \emph{rate stability} \cite{dai00}. A link $i$ is said to be rate stable if $\lim_{t\rightarrow\infty}A_i(t)/t=\lim_{n\rightarrow\infty}D_i(t)/t, \text{ w.p.1}$, so that a throughput of $\lambda_i$ can be achieved at link $i$. The network is rate stable if all links in the network are rate stable. It has been shown \cite{tassiulas92} that the largest achievable stability region is the convex hull of all independent sets, i.e., $\Lambda_{\textsf{opt}}={\bf Co}(\ml M)$, where ${\bf Co}(\cdot)$ denotes convex hull, and $\ml M$ is the set of independent sets of the interference graph. For the suboptimal schedulers that we consider here, the exact stability regions are hard to obtain, as one needs to prove stability for \emph{all possible arrival processes} with the same average rates. Instead, lower bounds are often provided. For maximal scheduling, it has been shown \cite{chaporkar08} that the following stability region can be achieved by any maximal scheduler:
\begin{equation}
  \Lambda_{\textsf{wc}}=\{{\bl \lambda}\in \mathbb{R}^N_+: \lambda_i+\sum_{j\in \ml{N}_i} \lambda_j\leq 1, \forall i\in \ml{V}_I\}
\end{equation}
where $\ml N_i=\{j\in\ml V_I: (i, j)\in\ml E_I\}$ is the set of neighbors of link $i$. As a comparison, we will prove later in this paper that a maximal scheduler with priority $\bl p$ can achieve the following stability region:
\begin{equation}\label{eqn_ap}
  \Lambda_{\bl p}=\{{\bl \lambda}\in \mathbb{R}^n_+: \lambda_i+\sum_{j\in \ml N_i} \lambda_j\bl 1_{\{p_i>p_j\}}\leq 1, \forall i\in \ml{V}_I\}
\end{equation}
where $\bl 1_{\{\cdot\}}$ is the indicator function, i.e., $\bl 1_{\{\text{true}\}}=1$, $\bl 1_{\{\text{false}\}}=0$. One can easily see that $\Lambda_{\textsf{wc}}\subseteq\Lambda_{\bl p}$ for any priority vector $\bl p$, so that $\Lambda_{\textsf{wc}}\subseteq \cap_{\bl p\in\ml P}\Lambda_{\bl p}$. In the following sections, we will propose a low complexity priority assignment scheme which can achieve $ \Lambda_{\textsf{sp}}=\cup_{\bl p\in\ml P} \Lambda_{\bl p}$.

\subsubsection{Scheduling Efficiency}

Another performance metric for a suboptimal scheduler is its \emph{scheduling efficiency}, which is defined as $\gamma=\sup\{\theta: \theta\Lambda_{\textsf{opt}}\subseteq \Lambda_{\pi}\}$, where $\Lambda_{\pi}$ is the stability region associated with the scheduler $\pi$. Thus, $\gamma$ denotes the largest achievable fraction of $\ml{A}_{\textsf{opt}}$ by the scheduler. In particular, $\gamma=1$ if the scheduling is optimal. It has been shown \cite{chaporkar08} that the worst case maximal scheduler has $\gamma_{\textsf{wc}}\geq 1/\Delta_{\tsf{wc}}$, where $\Delta_{\tsf{wc}}=\max_{i\in \ml{V}_I} \Delta_i$ is defined as the `interference degree' of the network, and $\Delta_i$ is the cardinality of the largest independent set in the subgraph induced by the links $\{i\}\cup\ml N_i$. For example, in the star shaped network in Fig. \ref{fig_star}, the interference degree of link 1 is $\Delta_1=8$, since there are at most 8 independent links in the subgraph induced by $\{1\}\cup \ml N_1$. Similarly, $\Delta_2=1$. It is easy to see that $\Delta_{\tsf{wc}}=8$, and therefore the worst case maximal scheduling can guarantee a scheduling efficiency of $1/8$.

For maximal scheduling with priority vector $\bl p$, we will show that its scheduling efficiency is $\gamma_{\bl p}\geq 1/\Delta^{\bl p}$, where $\Delta^{\bl p}=\max_{i\in \ml{V}_I}\Delta^{\bl p}_i$ is the \emph{prioritized} interference degree associated with priority $\bl p$. That is, $\Delta_i^{\bl p}$ is the cardinality of the largest independent set in the subgraph induced by links $\{i\}\cup\ml S^{\bl p}_i$. Note that $\Delta_i^{\bl p}\leq \Delta_i$ for any priority $\bl p$ and any link $i$, and therefore, the scheduling efficiency of the prioritized scheme is always better than the worst case. In Section \ref{sec_assignment}, we will show that a simple priority assignment scheme with the maximal scheduling can achieve a scheduling efficiency of $\gamma_{\textsf{sp}}\geq 1/\Delta_{\textsf{sp}}$, where $\Delta_{\textsf{sp}}=\min_{\bl p\in\ml P} \Delta^{\bl p}$.

\section{Maximal Scheduling with Fixed Priorities}
\label{sec_fixed}

In this section, we analyze the performance of maximal scheduling assuming a fixed priority $\bl p$ is always used. The optimal priority assignment will be presented in the next section. We begin by obtaining the stability region associated with a fixed priority vector $\bl p$.

\subsection{Lower Bound Stability Region}

We first show that the stability region in (\ref{eqn_ap}) is a lower bound for maximal scheduling with $\bl p$.

\begin{theorem}
  If ${\bl a}\in\Lambda_{\bl p}$, the network is rate stable under maximal scheduling with priority $\bl p$.
  \label{theorem_ap}
\end{theorem}

We first discuss the intuition behind the proof. According to Lemma \ref{lem_pms}, for any back-logged link $i$, maximal scheduling with priority $\bl p$ will schedule at least one packet departure from $\{i\}\cup \ml S_i^{\bl p}$. On the other hand, the definition of $\Lambda_{\bl p}$ in (\ref{eqn_ap}) implies that the total arrival rate in $\{i\}\cup \ml S_i^{\bl p}$ is no more than 1. Thus, \emph{assuming link $i$ always has packets}, the average total departure from the set $\{i\}\cup \ml S_i^{\bl p}$ is no smaller than the average total arrival, from which the stability result may follow.  However, rigorous proof is needed to show that the stability always holds \emph{without assuming that link $i$ is back-logged}, under general stochastic arrivals.

\begin{IEEEproof}
  We prove network stability using fluid limits \cite{dai00}, which is a general framework to analyze stochastic queueing systems. A short introduction of fluid limits is provided in Appendix \ref{apdx_fluid}. The proof of the theorem is in Appendix \ref{apdx_ap}.
\end{IEEEproof}

Having proved that $\Lambda_{\bl p}$ is a lower bound stability region, we next show its tightness result.

\begin{theorem}
  \label{theorem_ap_tight}
  For any network, if $\Lambda_{\bl p}\neq \Lambda_{\textsf{opt}}$, there exists an arrival rate vector $\bl \lambda\in \Lambda_{\textsf{opt}}$, which is arbitrarily close to $\Lambda_{\bl p}$, and a packet arrival process with average rate $\bl \lambda$, such that the network is unstable under maximal scheduling with priority $\bl p$.
\end{theorem}

\begin{IEEEproof}
  See in Appendix \ref{apdx_ap_tight}.
\end{IEEEproof}

Thus, the lower bound stability region $\Lambda_{\bl p}$ is tight. In Section \ref{sec_assignment}, we will show that the union of such regions $\Lambda_{\textsf{sp}}=\cup_{\bl p\in\ml P} \Lambda_{\bl p}$ can be achieved by a priority assignment algorithm. Before we do that, we analyze the scheduling efficiency represented by the lower bound region $\Lambda_{\bl p}$.

\subsection{Scheduling Efficiency}

We next show that the scheduling efficiency associated with $\Lambda_{\bl p}$ can be lower bounded by $\gamma_{\bl p}\geq 1/\Delta^{\bl p}$.

\begin{theorem}
  For any $\bl \lambda\in\Lambda_{\textsf{opt}}$, we have $(1/ \Delta^{\bl p})\bl \lambda\in \Lambda_{\bl p}$.
  \label{theorem_ap_approx_ratio}
\end{theorem}

\begin{IEEEproof}
  See in Appendix \ref{apdx_ap_approx_ratio}.
\end{IEEEproof}

We continue to show that $1/\Delta_{\textsf{sp}}$ is a lower bound on the scheduling efficiency of $\Lambda_{\tsf{sp}}$.

\begin{corollary}
  For any $\bl \lambda\in \Lambda_{\textsf{opt}}$, we have $(1/ \Delta_{\textsf{sp}})\bl \lambda\in\Lambda_{\textsf{sp}}$.
  \label{cor_ratio}
\end{corollary}

\begin{IEEEproof}
  See in Appendix \ref{apdx_ratio}.
\end{IEEEproof}

Compared to the interference degree $\Delta_{\tsf{wc}}$, the \emph{prioritized} interference degree $\Delta_{\textsf{sp}}$ can be much smaller, thereby achieving much larger scheduling efficiency. For example, in Fig. \ref{fig_star} (b), the interference degree is $\Delta_{\tsf{wc}}=8$, since link 1 has 8 independent neighbors. On the other hand, the prioritized interference degree implies that $\Delta_{\textsf{sp}}=1$ \emph{(globally optimal)}, which is achieved by assigning link 1 the highest priority. In fact, we have the following result:
\begin{corollary}
  For any wireless network with acyclic interference graph, we have $\Delta_{\textsf{sp}}=1$, so that the prioritized maximal scheduling is optimal.
  \label{cor_opt}
\end{corollary}

\begin{IEEEproof}
  See in Appendix \ref{adpx_cor_opt}.
\end{IEEEproof}

\subsection{Distributed Implementation}

We next briefly discuss the distributed implementation of proposed scheduling scheme. As a specific type of maximal scheduling, the prioritized maximal scheduling allows easy and distributed implementation in wireless networks. A direct implementation of the priority mechanism is as follows. We partition each time slot into a contention period and a data transmission period. Given a wireless network with $M$ priorities, we can divide the contention period of each time slot into $M$ mini-slots, such that a back-logged link with priority $k$ will back-off until the end of $k$-th mini-slot, and schedule itself for transmission if it does not hear any transmission intent (such as RTS) within its neighborhood. Otherwise link $k$ will wait until the next time slot. Note that it is possible to reduce the number of contention mini-time slot by introducing randomized back-off, such as \cite{lin09}, in which case the priority mechanism is implemented in a `soft' manner, to further reduce the scheduling overhead. It is also possible implement the prioritized maximal scheduling asynchronously by assigning back-off parameters such as the IFS in the 802.11 protocols. Such implementation topics are beyond the scope of this paper, and will be addressed in future research.

Summarizing the results in this section, we conclude that we can achieve a lower bound region of $\Lambda_{\tsf{sp}}=\cup_{\bl p\in\ml P} \Lambda_{\bl p}$ and scheduling efficiency bound of $1/\Delta_{\tsf{sp}}$ using maximal scheduling with static priorities, which is easily implemented in a distributed fashion. On the other hand, such performance guarantees hold only if \emph{the optimal priority is used}. In order to maximize the throughput performance, the priority vector $\bl p$ has to be carefully chosen. At the first glance, such priority assignment seems to be quite hard, since it involves an optimization over a set of $N!$ priority vectors. In the next section, we show that, somewhat surprisingly, the structure of $\Lambda_{\tsf{sp}}$ allows one to obtain the optimal priority very efficiently.

\section{Priority Assignment}
\label{sec_assignment}

In this section, we propose an online priority assignment algorithm to achieve $\Lambda_{\tsf{sp}}$. For the simplicity of exposition, we first consider a simple off-line case, where the estimated packet arrival rate $\hat{\bl \lambda}$ is fixed, and show that the priority assignment algorithm can output a stabilizing priority as long as $\hat{\bl \lambda}\in\Lambda_{\tsf{sp}}$. Later we will prove stability in the online case, where arrival rates must be estimated from general stochastic packet sources.

\subsection{An Offline Assignment}

The priority assignment algorithm is shown in Algorithm \ref{alg_spa}. At each step, the algorithm chooses a link $k$ with the smallest `total neighborhood arrival rate' $\hat{\lambda}_k+\sum_{j\in \ml{N}_k'}\hat{\lambda}_j$ in the \emph{reduced} interference graph $\ml{G}_I'$, and assigns it the lowest priority that is available \emph{locally}, i.e., link $k$ only needs to have higher priority than the neighboring links which are already removed. The algorithm then removes $k$ from $\ml{G}_I'$ and repeats. We next show that Algorithm \ref{alg_spa} implicitly solves a min-max optimization problem:

\begin{algorithm}[t]
  \caption{Local Priority Assignment}
  \begin{algorithmic}[1]
    \STATE {\bf Initialize:} $\ml V_I'\gets\ml V_I$, $\ml E_I'=\ml E_I$;
    \WHILE{$\ml V_I'\neq \emptyset$}
    \STATE Choose link $k$ such that
    \begin{equation}
      k=\arg\min_{i\in \ml{V}_I'} \{\hat{\lambda}_i+\sum_{j\in \ml{N}_i'}\hat{\lambda}_j\}
      \label{eqn_smin}
    \end{equation}
    \STATE If no neighbor of link $k$ has been removed, $p_k\gets N$. Otherwise $p_k\gets \beta-1$, where $\beta$ is the lowest priority among the removed neighbors of link $k$.
    \STATE Removed link $k$ from $\ml V_I'$ and its incident edges from $\ml E_I'$.
    \ENDWHILE
    \RETURN $\bl p$
  \end{algorithmic}
  \label{alg_spa}
\end{algorithm}

\begin{theorem}
  \label{theorem_opt}
  The priority vector $\bl p$ returned by Algorithm \ref{alg_spa} solves the following:
  \begin{equation}\label{eqn_argmin}
    \bl p\in\arg\min_{\bl p'\in\ml P}\max_{i\in \ml V_I} (\hat{\lambda}_i+\sum_{j\in\ml S^{\bl p'}_i}\hat{\lambda}_j).
  \end{equation}
\end{theorem}

\begin{IEEEproof}
  This can be proved using induction. See Appendix \ref{apdx_opt}.
\end{IEEEproof}

As an application of Theorem \ref{theorem_opt}, we next prove that Algorithm \ref{alg_spa} can achieve $\Lambda_{\tsf{sp}}$.

\begin{theorem}
  If $\hat{\bl \lambda}\in\Lambda_{\tsf{sp}}$, Algorithm \ref{alg_spa} will output a priority vector $\bl p$ such that $\hat{\bl \lambda}\in \Lambda_{{\bl p}}$.
  \label{theorem_spaopt}
\end{theorem}

\begin{IEEEproof}
  See Appendix \ref{apdx_spaopt}.
\end{IEEEproof}

Thus, we conclude that if $\hat{\bl \lambda}$ is fixed, Algorithm \ref{alg_spa} can output a stabilizing priority $\bl p$ as long as $\hat{\bl \lambda}\in\Lambda_{\tsf{sp}}$. Note that one important feature of Algorithm \ref{alg_spa} is that the priority can be \emph{reused}, as it is locally assigned. This can achieve a significant reduction in the total number of priorities and scheduling overhead. For example, the star network in Fig. \ref{fig_star} (b) only requires 2 distinct priorities. Further, for general wireless networks, one can often get a small number of priority levels by trading off a certain fraction of stability region. In fact, one can easily show that Algorithm \ref{alg_spa} only needs at most $d+1$ levels of priorities to achieve the worst case maximal scheduling performance bounds $\Lambda_{\tsf{wc}}$ and $\gamma_{\tsf{wc}}$, where $d$ is the maximum degree of the interference graph. The detailed analysis of such trade-off, on the other hand, is beyond the scope of this paper, and will be addressed in future research.

\subsection{Online Priority Assignment}

Having demonstrated the optimality of the offline priority assignment, we next implement it online with estimated arrival rates from stochastic packet arrival processes, and prove that the same optimality result still holds. To start with, we partition time into frames, each of which has length $T$. Throughout each frame $l$, a fixed priority $\bl p(l)$ is used, which is computed as follows. For the first frame, we assume $\bl p(1)$ is arbitrary. At the beginning of each subsequent frame, $\bl p(l)=\bl p(l-1)$ if the estimated arrival rate satisfies $\hat{\bl \lambda}(l-1)\in \Lambda_{\bl p(l-1)}$, where 
\begin{equation}
  \hat{\bl \lambda}(l-1)=\bl A((l-1)T)/(l-1)T
\end{equation}
Otherwise we set $\bl p(l)=\bl p$, where $\bl p$ is returned by Algorithm \ref{alg_spa} with $\hat{\bl \lambda}(l-1)$. We next show network stability in the following theorem:

\begin{theorem}
  The network is rate stable under the online priority assignment scheme if $\bl \lambda\in{\bf int}(\Lambda_{\tsf{sp}})$, where ${\bf int}(\cdot)$ denotes the interior.
  \label{theorem_dynamic}
\end{theorem}


\begin{IEEEproof}
  It is sufficient to prove that the priority process $\{\bl p(l)\}$ will `converge' to a feasible one in a finite number of frames. The details are in Appendix \ref{apdx_opt}.
\end{IEEEproof}


\section{Simulation Results}
\label{sec_simulation}

In this section, we evaluate the performance of the proposed priority scheduling scheme by MATLAB simulation. All the simulation results in this section are obtained from 30 independent simulations over a period of $10^5$ time slots. The packet arrival processes are i.i.d and independent across different links.

\subsection{Intersecting Cliques}

 \begin{figure}[!t]
  \includegraphics[width=3.6in]{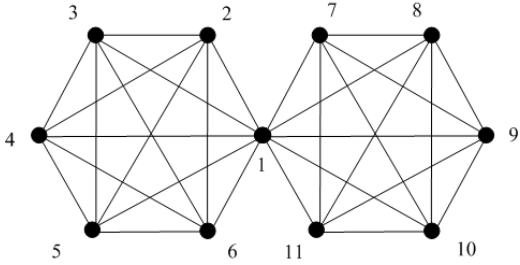}
  \caption{\label{fig_2_clique} An interference graph consisting of
two cliques of size 6 each, with one common link 1.}
\end{figure}

\begin{figure}[!t]
\centering
\includegraphics[width=3.5in]{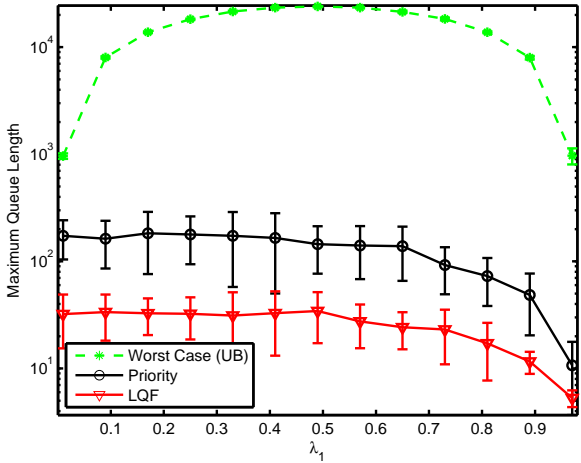}
\caption{The performance comparison of different scheduling
algorithms in the 2-clique network.} \label{fig_2_clique_sim}
\end{figure}

We first consider a wireless network with 11 links as shown in Fig. \ref{fig_2_clique}, where the center link 1 is at the intersection of two cliques. Thus, link 1 interfere with both local clusters, and is the bottleneck of the network. We assume that every link other than link 1 has an arrival rate of $(0.99-\lambda_1)/5$, so that each clique has a total arrival rate of $0.99$. We simulate three types of scheduling algorithms: 1) a maximal scheduler with a suboptimal priority vector, as an upper bound on the worst-case throughput performance of maximal scheduling, 2) maximal scheduling with the optimal priority vector obtained by the online priority assignment algorithm, and 3) the LQF scheduling. For the prioritized maximal scheduling, we choose $T=100$. The priority converges after a few time frames as the arrival processes are i.i.d. Fig. \ref{fig_2_clique_sim} shows the maximum queue lengths under different values of $\lambda_1$ over $10^5$ time slots simulation, along with $95\%$ confidence intervals.

\subsubsection{Throughput Optimality}

The network is unstable under the worst-case maximal scheduling, which can be clearly observed by the significant larger queue lengths than the other two schedulers (note that $10^5$ time slots were simulated). On the other hand, the network is always stable under maximal scheduler with the optimal priority. In fact, for this topology, the optimal priority scheduling scheme is \emph{globally optimal}, since one can easily verify that $\gamma_{\tsf{sp}}=1$. Thus, we can obtain significant throughput improvement by properly optimizing the priorities.

\subsubsection{LQF Scheduling}

The network is also stable under LQF scheduling. In fact, it can be shown that LQF scheduling is throughput optimal for such topology, due to the `local pooling' condition \cite{dimakis06}. It has also been widely observed that the LQF scheduling can achieve good throughput performance in general wireless networks, at the expense of frequent update of priorities, which may incur large scheduling overhead. Compared to LQF scheduling, the static priority based maximal scheduling can achieve similar throughput performance, with smaller scheduling overhead.

\subsection{Random Topology}

\begin{figure}[!t]
  \centering
  \includegraphics[width=3.5in]{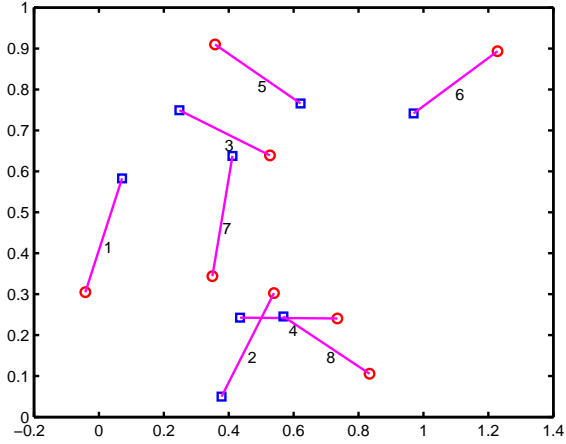}
  \caption{A random wireless network with 8 links, where square nodes are the transmitters, and round nodes are the receivers.}
  \label{fig_random_network}
\end{figure}

\begin{figure}[!t]
\centering
\includegraphics[width=3.5in]{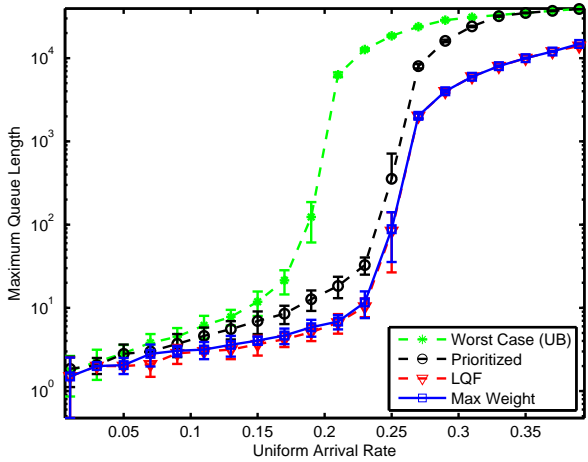}
\caption{The simulation result in the random network with 8 links,
where the maximum queue lengths are shown under uniform arrival
rates.} \label{fig_random_sim}
\end{figure}

We next consider a random wireless network with 8 links, whose communication graph is shown in Fig. \ref{fig_random_network}. In the figure, the square nodes are the transmitters, and the round nodes are the receivers. The interference graph is constructed by placing a guard zone \cite{hasan07} around the receiver of each link, so that two links form an edge if one's transmitter is inside the guard zone associated with the other. We consider all the scheduling algorithms in the previous case, and the optimal max-weight scheduling \cite{tassiulas92} as a benchmark. The simulation results are shown in Fig. \ref{fig_random_sim}, where the maximum queue lengths are shown after $10^5$ time slots, with $95\%$ confidence intervals. Similar convergence results are observed for the online priorities, due to the fast convergence of the arrival processes.

\subsubsection{Maximal Scheduling}

The network is unstable under the worst-case maximal scheduling with arrival rate above $0.2$. From the queue lengths of the max weight scheduling, it can be observed that the boundary of the stability region is around $0.25$. Thus, for this random network, the worst case maximal scheduling can achieve a scheduling efficiency of no more than around $0.8$.

\subsubsection{Prioritized Maximal Scheduling}

Maximal scheduling with optimal priority achieves the same maximum uniform throughput as the max-weight scheduling, although with larger queue lengths. Thus, compared to the worst-case maximal scheduling, we conclude that we can achieve significant throughput improvement by choosing the priority vectors carefully. Further, compared to the max-weight scheduling, the optimal priority based maximal scheduling can achieve the same throughput with low complexity.

\subsubsection{LQF Scheduling}

The LQF scheduling also achieves the network stability for all arrival rates, with smaller queue lengths than the optimal prioritized maximal scheduling. However, note that this is achieved at the expense of more control overhead associated with frequent priority updates, due to the changes in the queue lengths, which induces global changes to the priorities in each time slot. On the other hand, the priority updates of our scheduler with the estimated arrival rates happen very infrequently, and therefore requires much smaller scheduling overhead.

\section{Conclusion}
\label{sec_conclusion}

In this paper, we proposed a static priority based scheme to improve the performance of maximal scheduling in wireless networks. The scheduling has low complexity, and is easily amendable for distributed implementation. We first formulated a tight lower bound stability region for maximal scheduling with a fixed priority, and then discussed its scheduling efficiency. We next introduced an online priority assignment scheme, which can compute the optimal priority based on estimated packet arrival rates. Future research will focus on the trade-off between the throughput performance and the scheduling overhead, such as the number of distinct priorities.

\appendices

\section{Proof of Lemma \ref{lem_pms}}
\label{apdx_pms}
\begin{IEEEproof}
  According to the scheduler, the links in $\ml{S}_i^{\bl p}$ are always considered before link $i$. Thus, when a back-logged link $i$ is being considered by the scheduler, either there is already a scheduled link in $\ml{S}_i^{\bl p}$, or link $i$ is put to the schedule. In both cases, there is at least one packet departure among the links in $\{i\}\cup \ml{S}_i^{\bl p}$, and therefore, the lemma follows.
\end{IEEEproof}

\section{Fluid Limits}
\label{apdx_fluid}

In this section we briefly introduce fluid limits, which is a general framework to analyze stochastic queueing systems. For details, please refer to \cite{dai00} and the references therein.

\subsection{Construction of Fluid Limits}

Given the network dynamics $({\bl Q}(t), {\bl A}(t), {\bl D}(t))_{t=0}^\infty$, we first extend the support from $\mathbb{N}$ to $\mathbb{R}_+$ using linear interpolation. For a fixed sample path $\omega$, define the following fluid scaling 
\begin{equation}
  f^r(t, \omega)={f(rt, \omega)/r}
\end{equation}
where the function $f(\cdot)$ can be ${Q}_i(\cdot), {A}_i(\cdot)$ or ${D}_i(\cdot)$. It can be verified that these functions are uniformly Lipschitz-continuous, i.e., for any $t>0$ and $\delta>0$, we have 
\begin{equation}
  f^{r}(t+\delta)-f^{r}(t)\leq K\delta
\end{equation}
where the positive constant $K$ is $A_{\max}$ for functions $A_i(\cdot)$ and $Q_i(\cdot)$ and 1 for functions $D_i(\cdot)$. Thus, these functions are equi-continuous. According to the Arz\'ela-Ascoli Theorem \cite{royden88}, any sequence of equi-continuous functions $\{f^{r_n}(t)\}_{n=1}^\infty$ contains a subsequence $\{f^{r_{n_k}}(t)\}_{k=1}^\infty$, such that 
\begin{equation}
\lim_{k\rightarrow\infty}\sup_{\tau\in[0,t]}|f^{r_{n_k}}(\tau)-\bar{f}(\tau)|=0  
\end{equation}
with probability 1, where $\bar{f}(t)$ is a uniformly continuous function (and therefore differentiable almost everywhere \cite{royden88}). Define any such limit $(\bar{\bl Q}(t), \bar{\bl A}(t), \bar{\bl D}(t))$ as a fluid limit.

\subsection{Properties of Fluid Limits}

The following properties holds for any fluid limit
\begin{eqnarray}
  \bar{A}_i(t)&=&\lambda_it\label{eqn_ai_bar}\\
  \diff \bar{Q}(t) &=&0\quad\textrm{ if }\bar{Q}(t)=0
  \label{eqn_q_bar}
\end{eqnarray}
where (\ref{eqn_ai_bar}) is because of (\ref{eqn_slln}) and the functional SLLN, and (\ref{eqn_q_bar}) is because any regular point $t$ with $\bar{Q}(t)=0$ achieves the minimum value (since $\bar{Q}(t)\geq 0$), and therefore has zero derivative. We further have the following lemma, which provides a sufficient condition about rate stability \cite{dai00}:
\begin{lemma}
  \label{lem_fluid_stable}
  The network is rate stable if any fluid limit with $\bar{\bl Q}(0)={\bf 0}$ has $\bar{\bl Q}(t)={\bf 0}, \forall t\geq 0$.
\end{lemma}

\section{Proof of Theorem \ref{theorem_ap}}
\label{apdx_ap}

\begin{IEEEproof}
  Since the priority vector $\bl p$ is fixed, for ease of notation, we relabel the links in decreasing order of  priorities according to $\bl p$. Thus, link 1 has the highest priority, and link $N$ has the lowest priority. Fix a sample path $\omega$ where SLLN applies, and consider the following Lyapunov function 
  \begin{equation}
    L(t)=\sum_{i\in \ml{V}_I}{\bar Q}^2_i(t)  
  \end{equation}
  According to Lemma \ref{lem_fluid_stable}, it is sufficient to prove that $\dot{L}(t)\leq 0$ if $\bar{\bl Q}(0)={\bf 0}$, since we then have 
  \begin{equation}
    L(t)=L(0)+\int_{0}^t \dot{L}(\tau)d\tau\leq L(0) = 0  
  \end{equation}
  from which we conclude that $\bar{\bl Q}(t)=\bl 0$ for all $t\geq 0$. To prove this, in the following we will show that, by induction, $\diff{\bar{Q}}^2_{i}(t)\leq 0$ for each link $i$ if $\bar{\bl Q}(0)={\bf 0}$.

  We first consider the link $1$, which has the highest priority according to $\bl p$. Note that if $ \bar{Q}_{1}(t)=0$, we have 
  \begin{equation}
    \diff{\bar{Q}}^2_{1}(t)=2 \bar{Q}_{1}(t)\dot{\bar{Q}}_{1}(t)=0
  \end{equation}
  Now suppose that, on the contrary, $\bar{Q}_{1}(t)>0$ at some $t>0$. Then there exists a constant $\varepsilon>0$ such that $\bar{Q}_{1}(t)>\varepsilon>0$. Since $\bar{Q}_{1}(t)$ is uniformly continuous, there also exists $\delta>0$ such that 
  \begin{equation}
    {\bar Q}_{1}(\tau)>\varepsilon/2 \text{ for } \tau\in(t-\delta, t+\delta).
  \end{equation}
Now consider any converging subsequence $\{f^{r_{n_k}}(t)\}_{k=1}^\infty$ for the fluid limit. For sufficiently large $k$, we have 
\begin{equation}
  Q_{1}^{r_{n_k}}(\tau)>\varepsilon/4, \forall \tau\in(t-\delta, t+\delta)
\end{equation}
which implies that 
\begin{equation}
  Q_{1} (\tau)>r_{n_k}\varepsilon/4\geq 1, \forall \tau\in(r_{n_k}(t-\delta), r_{n_k}(t+\delta))  
\end{equation}
That is, link ${1}$ is always back-logged during the time interval $(r_{n_k}(t-\delta), r_{n_k}(t+\delta))$. Due to the prioritized maximal scheduling specification, link $1$ transmits in every time slot in this interval, since it has the highest priority. Thus, we conclude that 
\begin{equation}
  D_{1}(r_{n_k}(t+\delta))-D_{1}(r_{n_k}(t-\delta))=2r_{n_k}(t+\delta)
\end{equation}
After taking limit as $k\rightarrow\infty$ we have 
\begin{equation}
  {\bar{D}}_{1}(t+\delta)-{\bar{D}}_{1}(t-\delta)=2\delta
\end{equation}
which implies that $\dot{\bar{D}}_{1}(t)=1$ since $\delta>0$ can be arbitrarily small. Therefore, we conclude that 
\begin{eqnarray}
    \diff{\bar{Q}}^2_{1}(t)&=&2\bar{Q}_{1}(t)\dot{\bar{Q}}_{1} (t)\\
    &=&2\bar{Q}_{1}(t)(\lambda_i-\dot{\bar{D}}_{1}(t))\\
    &=&2\bar{Q}_{1}(t)(\lambda_i-1)\\
    &\leq& 0
\end{eqnarray}
where the last equality is due to the assumption that $\bl \lambda\in \Lambda_{\bl p}$. Thus, we have $\diff{\bar{Q}}^2_{1}(t)\leq 0$ and $\bar{Q}_{1}(t)=0 $ for all $ t\geq 0 $.

We next proceed by induction. Suppose that $\diff{\bar{Q}}^2_{k}(t)\leq 0$ and $ \bar{Q}_{k}(t)=0 $ for all $t\geq 0$ and $k\leq l-1$, i.e., the first $l-1$ highest priority links. Now consider the link $l$, which has the $l$-th highest priority. Note that if $\bar{Q}_{l}(t)=0$ we have $\dot{\bar{Q}}_{l}(t)=0 $. Now suppose $\bar{Q}_{l}(t)>0$ for some $t>0$. Following the same argument as for link $1$, we conclude that there is some interval $(r_{n_k}(t-\delta), r_{n_k}(t+\delta))$ during which $ Q_{l}(\tau) $ is nonempty. According to Lemma \ref{lem_pms}, in each time slot the maximal scheduler with priority $\bl p$ will schedule at least one departure in $\{l\}\cup\ml S^{\bl p}_{l}$, and therefore, we have
\begin{eqnarray*}
  &&\Big((D_{l}(r_{n_k}(t+\delta))+\sum_{j\in\ml S_{l}^{\bl p}}D_j(r_{n_k}(t+\delta))\Big)\\
  &&-\Big(D_{l}(r_{n_k}(t-\delta))+\sum_{j\in\ml S_{l}^{\bl p}}D_j(r_{n_k}(t-\delta))\Big)\geq 2r_{n_k}\delta
\end{eqnarray*}
which implies, after taking $k\rightarrow\infty$, that 
\begin{equation}
  \dot{\bar{D}}_{l}(t)+\sum_{j\in S_{l}^{\bl p}}\dot{\bar{D}}_j(t)\big)\geq 1
\end{equation}

Thus, we conclude that
\begin{eqnarray*}
\diff{\bar{Q}}^2_{l}(t)&\stackrel{(a)}{=} & 2{\bar{Q}}_{l}(t)(\dot{\bar{Q}}_{l}(t)+\sum_{j\in\ml S_{l}^{\bl p}} \dot{\bar{Q}}_j(t))\\
&=&2\bar{Q}_{l}(t)(\lambda_{l}+\sum_{j\in\ml S_{l}^{\bl p}}\lambda_j-\big(\dot{\bar{D}}_i(t)+\sum_{j\in \ml S_{l}^{\bl p}}\dot{\bar{D}}_j(t)\big))\\
&\leq& 2\bar{Q}_{l}(t)(\lambda_{l}+\sum_{j\in\ml S_{l}^{\bl p}}\lambda_j-1)\stackrel{(b)}{\leq} 0
\end{eqnarray*}
where $ (a) $ is because, by induction hypothesis, $ \bar{Q}_j(t)=0 $ for all $ t\geq 0 $ and all higher priority neighbors $ j\in\ml S_{l}^{\bl p}$, and $ (b) $ is because 
\begin{equation}
  \lambda_{l}+\sum_{j\in\ml S_{l}^{\bl p}}\lambda_j\leq 1
\end{equation}
since $\bl \lambda\in \Lambda_{\bl p} $. Thus, by induction, we conclude that $ \diff\bar{Q}^2_{i}(t)\leq 0 $ for all $ t\geq 0$ and all links in the network, from which the theorem follows.
\end{IEEEproof}

\section{Proof of Theorem \ref{theorem_ap_tight}}
\label{apdx_ap_tight}
\begin{IEEEproof}
    If $\Lambda_{\bl p}\neq \Lambda_{\textsf{opt}}$, there must be an arrival rate $\bl \lambda\in \Lambda_{\textsf{opt}}$ such that for some link $i$, we have $\lambda_i + \sum_{j\in\ml S^{\bl p}_i} \lambda_j >1$. Further, the links in $\{i\}\cup\ml S^{\bl p}_i$ can not form a clique, since in that case we will have $\bl \lambda\not\in \Lambda_{\textsf{opt}}$. Thus, we can always find two independent links $j$ and $k$ in the set $\ml S^{\bl p}_i$. Now consider the following arrival rates: $\lambda'_i=\varepsilon$, $\lambda'_j=\lambda'_k=1/2$, and $\lambda'_l=0$ for any other link $l$. It is easily seen that $\bl \lambda'\in \ml A_{\textsf{opt}}$, since one can simply alternate between the two schedules $\{i\}$ and $\{j, k\}$ in odd and even time slots to achieve network stability. Note that by adjusting the parameter $\varepsilon$, the arrival rate vector $\bl \lambda'$ can be arbitrarily close to $\Lambda_{\bl p}$. Now, we consider the following arrival process with arrival rate $\bl \lambda'$. In every odd time slot, a packet arrives at link $j$, and in every even time slot, a packet arrives at link $k$. Thus, according to the maximal scheduling with priority $\bl p$, these packets are immediately transmitted in the next time slot. Finally, in each time slot, a packet arrives at link $i$ independently with probability $\varepsilon$. Thus, link $i$ is never scheduled by the maximal scheduler, and is therefore starved.
\end{IEEEproof}

\section{Proof of Theorem \ref{theorem_ap_approx_ratio}}
\label{apdx_ap_approx_ratio}

\begin{IEEEproof}
   For any link $i$, according to the definition of $\Delta^{\bl p}_i$, there are at most $\Delta^{\bl p}_i$ packet departures among $\{i\}\cup{\ml S^{\bl p}_i}$ in each time slot, since the transmitting links must form an independent set in the subgraph induced by $\{i\}\cup{\ml S^{\bl p}_i}$. Thus, if the network is stable, the total average arrivals in $\{i\}\cup{\ml S^{\bl p}_i}$ must be no more than the total average departures, i.e., $\lambda_i+\sum_{j\in\ml S^{\bl p}_i}\lambda_j\leq \Delta^{\bl p}_i\leq \Delta^{\bl p}, \quad\forall i\in \ml V_I$. Multiplying both sides of the above inequality with $1/ \Delta^{\bl p}$, and recalling the definition of $\ml A_{\bl p}$ in (\ref{eqn_ap}), we conclude that $(1/ \Delta^{\bl p})\bl \lambda\in \Lambda_{\bl p}$ and the theorem follows.
\end{IEEEproof}

\section{Proof of Corollary \ref{cor_ratio}}
\label{apdx_ratio}
\begin{IEEEproof}
    Note that $\ml P$ is a finite set, and therefore there must be $\bl p^\star\in\ml P$, such that $\Delta^{\bl p^\star}=\Delta_{\tsf{sp}}=\min_{\bl p\in\ml P}\Delta^{\bl p}$. Thus, according to theorem \ref{theorem_ap_approx_ratio}, we have $(1/ \Delta_{\textsf{sp}})\bl \lambda=(1/ \Delta^{\bl p^\star})\bl \lambda\in \Lambda_{\bl p^\star}\subseteq \Lambda_{\tsf{sp}}$, from which the claim holds.
\end{IEEEproof}

\section{Proof of Corollary \ref{cor_opt}}
\label{adpx_cor_opt}
\begin{IEEEproof}
  Consider any connected component $\ml T$ of the interference graph $\ml G_I$. Since $\ml G_I$ is acyclic, $\ml T$ must be a tree. Pick an arbitrary node in $\ml T$ as the root. Now, consider the following priority assignment on the nodes in $\ml T$. We first assign the lowest priorities to all the leaves of $\ml T$ (the priorities within the leaves are arbitrary) and then remove them from $\ml T$. Next, we assign the next levels of the lowest priorities to the newly formed leaves, and then remove them. This procedure repeats until $\ml T$ is empty. It can be easily seen that the assigned priority $\bl p$ implies that each node $i\in\ml T$ will only have one neighbor with higher priority, which is its parent node. Therefore, we have $\Delta_i^{\bl p}=1$. Since the argument can be applied to the other connected components of $\ml G_I$, we conclude that $\Delta^{\bl p}=1$. Finally, we note that $1\leq \Delta_{\textsf{sp}}\leq \Delta^{\bl p}=1$, from which the claim follows.
\end{IEEEproof}

\section{Proof of Theorem \ref{theorem_opt}}
\label{apdx_opt}

\begin{IEEEproof}
  Let a priority $\bl p'\in \ml P$ be given. It is sufficient to prove that 
  \begin{equation}
      \hat{\lambda}_k+\sum_{j\in\ml S^{\bl p}_k}\hat{\lambda}_j \leq \max_{i\in \ml V_I} (\hat{\lambda}_i+\sum_{j\in\ml S^{\bl p'}_i}\hat{\lambda}_j)
  \end{equation}
for any link $k\in\ml V_I$. For notation simplicity, we relabel the links according to the reverse order of the priority $\bl p$, so that link 1 has the lowest priority, and link $N$ has the highest priority. Now consider the first iteration of Algorithm \ref{alg_spa}, and denote $1'$ as the lowest priority link according to ${\bl p'}$. We have
\begin{eqnarray*}
   \hat{\lambda}_{1}+\sum_{j\in \ml{N}_{1}'}\hat{\lambda}_j &\stackrel{(a)}{\leq}& \hat{\lambda}_{1'}+\sum_{j\in \ml{N}_{1'}'}\hat{\lambda}_j\\
   &\stackrel{(b)}{=}&\hat{\lambda}_{1'}+\sum_{j\in \ml{S}^{\bl p'}_{1'}}\hat{\lambda}_j\leq \max_{i\in \ml V_I}(\hat{\lambda}_i+\sum_{j\in\ml S^{\bl p'}_i}\hat{\lambda}_j)
\end{eqnarray*}
Note that here, the sets $\ml N_1'$ and $\ml N_{1'}'$ refer to the neighbors of link $1$ and $1'$ at the first iteration of Algorithm \ref{alg_spa}, respectively. $(a)$ is because of (\ref{eqn_smin}), and $(b)$ is because $\ml N_{1'}'=\ml S_{1'}^{\bl p'}$, since link $1'$ has the lowest priority according to $\bl p'$. Now consider the second iteration of Algorithm \ref{alg_spa}, with new reduced interference graph by removing link $1$. Similarly, denote $2'$ as the lowest priority link according to ${\bl p'}$ in the reduced interference graph $\ml G_I'$ at the second iteration of Algorithm \ref{alg_spa}. We have
\begin{eqnarray*}
  \hat{\lambda}_{2}+\sum_{j\in \ml{N}_{2}'}\hat{\lambda}_j &\leq& \hat{\lambda}_{2'}+\sum_{j\in \ml{N}_{2'}'}\hat{\lambda}_j\\
  &\stackrel{(a)}{\leq}&\hat{\lambda}_{2'}+\sum_{j\in \ml{S}^{\bl p'}_{2'}}\hat{\lambda}_j\leq\max_{i\in \ml V_I}(\hat{\lambda}_i+\sum_{j\in\ml S^{\bl p'}_i}\hat{\lambda}_j)
\end{eqnarray*}
where $(a)$ is $\ml S_{2'}^{\bl p'}$ refers to the original interference graph, whereas $\ml N_{2'}'$ refers to the higher priority neighbors in the reduced interference graph. Similarly, by repeating the above arguments, we conclude that 
\begin{equation}
  \hat{\lambda}_{i}+\sum_{j\in \ml{N}_{i}'}\hat{\lambda}_j\leq \max_{i\in \ml V_I}(\hat{\lambda}_i+\sum_{j\in\ml S^{\bl p'}_i}\hat{\lambda}_j)
\end{equation}
for each iteration of $i$ of the Algorithm \ref{alg_spa}. Finally, according to Algorithm \ref{alg_spa}, the links removed later are always assigned higher priorities. Therefore, we have $\ml S_{i}^{\bl p}=\ml N_{i}'$, which implies that 
\begin{eqnarray}
  \hat{\lambda}_{i}+\sum_{j\in \ml{S}^{\bl p}_{i}}\hat{\lambda}_j&=&\hat{\lambda}_{i}+\sum_{j\in \ml{N}_{i}'}\hat{\lambda}_j\\
  &\leq& \max_{i\in \ml V_I}(\hat{\lambda}_i+\sum_{j\in\ml S^{\bl p'}_i}\hat{\lambda}_j)
\end{eqnarray}
for all $i\in\ml V_I$, from which the theorem follows.
\end{IEEEproof}

\section{Proof of Theorem \ref{theorem_spaopt}}
\label{apdx_spaopt}
\begin{IEEEproof}
   Since $\hat{\bl \lambda}\in\Lambda_{\tsf{sp}}$, there is $\bl p'\in\ml P$ such that $\hat{\bl \lambda}\in\Lambda_{\bl p'}$, which implies that $\max_{i\in \ml V_I}(\hat{\lambda}_i+\sum_{j\in\ml S^{\bl p'}_i}\hat{\lambda}_j)\leq 1$. From Theorem \ref{theorem_opt}, Algorithm \ref{alg_spa} will return a priority $\bl p$ such that 
   \begin{eqnarray}
  \max_{i\in \ml V_I} (\hat{\lambda}_i+\sum_{j\in\ml S^{\bl p}_i}\hat{\lambda}_j)&\leq& \max_{i\in \ml V_I} (\hat{\lambda}_i+\sum_{j\in\ml S^{\bl p'}_i}\hat{\lambda}_j)\\
  &\leq& 1
   \end{eqnarray}
from which we conclude that $\hat{\bl \lambda}\in \Lambda_{\bl p}$. Therefore, the theorem follows.
\end{IEEEproof}

\section{Proof of Theorem \ref{theorem_dynamic}}

\begin{IEEEproof}
  We partition the set of priority vectors into three disjoint subsets: 
  \begin{eqnarray*}
  \ml P=\ml P_1\cup \ml P_2\cup \ml P_3,
  \end{eqnarray*}
such that ${\bl \lambda}\in\cap_{\bl p\in\ml P_1}{\bf int}(\Lambda_{\bl p})$, ${\bl \lambda}\in\cap_{\bl p\in\ml P_2}{\bf bd}({\Lambda}_{\bl p})$, and ${\bl \lambda}\in\cap_{\bl p\in\ml P_3}{\Lambda}^c_{\bl p}$, where ${\bf int}(\cdot)$ denotes the interior, ${\bf bd}(\cdot)$ denotes the boundary, and $(\cdot)^c$ denotes the complement. Thus, ${\bl \lambda}$ is `strictly' stable for any priority from $\ml P_1$, and is `critically' stable for any priority from $\ml P_2$, but is unstable under any priority from $\ml P_3$. In the following, we will show that after a finite number of frames, the sequence of priority vectors $\{\hat{\bl p}(l)\}$ will stay fixed at a priority vector in either $\ml P_1$ or $\ml P_2$. Thus, an identical argument using fluid limits as shown in the proof of Theorem \ref{theorem_ap} can be applied to show that the network is stable.

First, since ${\bl \lambda}\in\cap_{\bl p\in\ml P_1}{\bf int}(\Lambda_{\bl p})$, there exists an $\varepsilon_1>0$ such that, for any $\hat{\bl \lambda}$ satisfying $\|\hat{\bl \lambda}-\bl \lambda\|_2<\varepsilon_1$, we have $\hat{\bl \lambda}\in\cap_{\bl p\in\ml P_1}{\bf int}(\Lambda_{\bl p})$. Further, since $\bl \lambda$ is `critically' stable under any priority in $\ml P_2$, we can choose $\varepsilon_2>0$ such that for any $\hat{\bl \lambda}$ satisfying $\|\hat{\bl \lambda}-\bl \lambda\|_2<\varepsilon_2$, and any $\bl p\in \ml P_1$, $\bl p'\in\ml P_2$, we have 
\begin{eqnarray*}
  \max_{i\in \ml V_I} (\hat{\lambda}_i+\sum_{j\in\ml S^{\bl p}_i}\hat{\lambda}_j)<\max_{i\in \ml V_I} (\hat{\lambda}_i+\sum_{j\in\ml S^{\bl p'}_i}\hat{\lambda}_j)
\end{eqnarray*}
Thus, if Algorithm \ref{alg_spa} is executed with the above $\hat{\bl \lambda}$, the output priority vector must lie in $\ml P_1$, according to Theorem \ref{theorem_opt}. Finally, note that $\cap_{\bl p\in\ml P_3}{\Lambda}^c_{\bl p}$ is an open set, we can choose $\varepsilon_3>0$ sufficiently small, such that any $\hat{\bl \lambda}$ satisfying $\|\hat{\bl \lambda}-\bl \lambda\|_2<\varepsilon_3$ still satisfies $\hat{\bl \lambda}\in\cap_{\bl p\in\ml P_3}{\Lambda}^c_{\bl p}$. Now, we choose $\varepsilon'=\min(\varepsilon_1, \varepsilon_2, \varepsilon_3)$, and because of the SLLN in (\ref{eqn_slln}), we can choose $L$ to be large enough such that for any $l>L$, we have $\|\hat{\bl \lambda}(l)-\bl \lambda\|_2<\varepsilon'$. Thus, if Algorithm \ref{alg_spa} is executed for any $l>L$, we have $\bl p(l)\in\ml P_1$, because of (\ref{eqn_argmin}). Further, for any $l>L$, if Algorithm \ref{alg_spa} is executed, the priority vector will stay at the output result $\bl p\in\ml P_1$, since by assumption, $\hat{\bl \lambda}(l)\in \Lambda_{\bl p}$. Finally, we only need to consider the case where Algorithm \ref{alg_spa} is not executed for all $l\geq L$. It is clear that in such case, $\bl p(l)\not\in \ml P_3$ for any $l\geq L$. Thus, for sufficiently large $l$, the priority vector stays at a point in either $\ml P_1$ or $\ml P_2$ without invoking Algorithm \ref{alg_spa}, from which we can conclude that the network is stable.
\end{IEEEproof}

\bibliographystyle{IEEEtran}
\bibliography{IEEEabrv,wireless}

\begin{IEEEbiography}[{\includegraphics[width=1in,height=1.25in,clip,keepaspectratio]{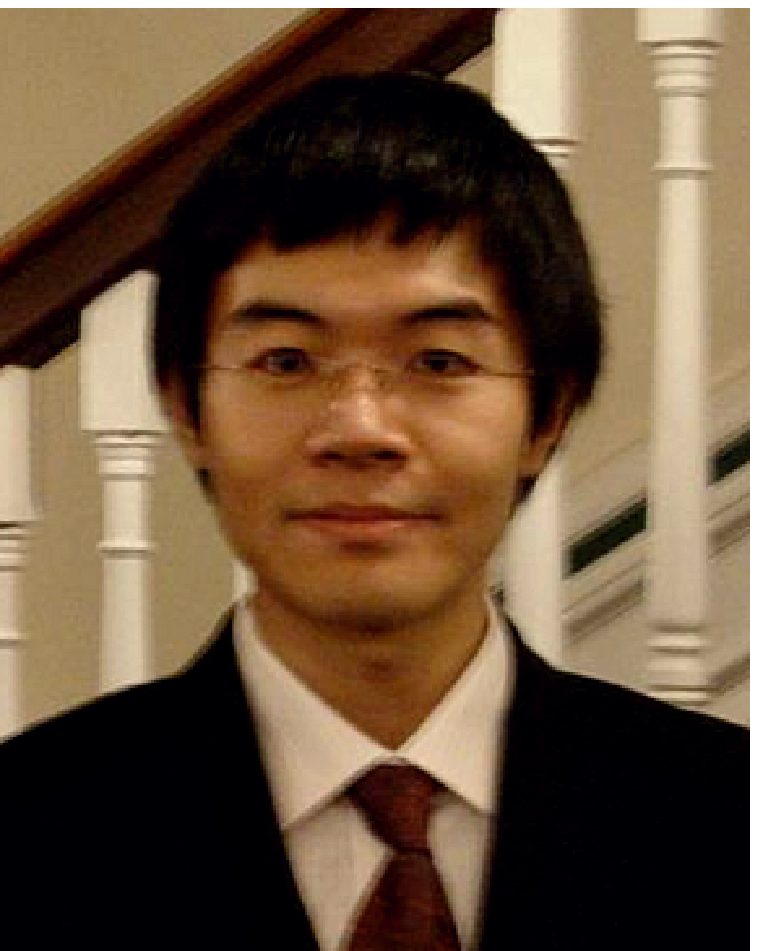}}]{Qiao Li} (S'07) received the B.Engg. degree from the Department of Electronics Information Engineering, Tsinghua University, Beijing China, in 2006. He received the M.S. degree from the Department of Electrical and Computer Engineering, Carnegie Mellon University, Pittsburgh, PA USA, in 2008. 
  
  He is currently a Ph.D. Candidate in the Department of Electrical and Computer Engineering, Carnegie Mellon University. His research interests include distributed algorithms, smart grid technologies, and wireless networking.
\end{IEEEbiography}

\begin{IEEEbiography}[{\includegraphics[width=1in,height=1.25in,clip, keepaspectratio]{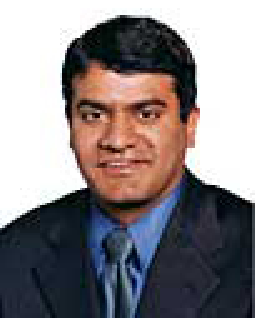}}]{Rohit Negi} (S'98-M'00) received the B.Tech. degree in electrical engineering from the Indian Institute of Technology, Bombay, in 1995. He received the M.S. and Ph.D. degrees from Stanford University, CA, in 1996 and 2000, respectively, both in electrical engineering. 
  
  Since 2000, he has been with the Electrical and Computer Engineering Department, Carnegie Mellon University, Pittsburgh, PA, where he is a Professor. His research interests include signal processing, coding for communications systems, information theory, networking, cross-layer optimization, and sensor networks. Dr. Negi received the President of India Gold Medal in 1995. 
\end{IEEEbiography}
\end{document}